\documentclass[10pt]{article}
\usepackage[OE]{express}
\usepackage{soul}
\newcommand{\dC}{$^{\rm o}$C}
\newcommand{\ms}{{m s$^{-1}$}}
\newcommand{\fabper}{Fabry-P{\'e}rot}

\begin{document}

\title{Frequency stability characterization of a broadband fiber {\fabper} interferometer}

\author{Jeff Jennings,\authormark{1,2,8} Samuel Halverson,\authormark{3,4,5,6,7} Ryan Terrien,\authormark{1} Suvrath Mahadevan,\authormark{3,4,5} Gabriel Ycas,\authormark{1} and Scott A. Diddams\authormark{1,2,9}}

\address{\authormark{1}National Institute of Standards and Technology, Boulder, CO 80305, USA\\
\authormark{2}Department of Physics, University of Colorado, Boulder, CO 80309, USA\\
\authormark{3}Department of Astronomy \& Astrophysics, The Pennsylvania State University, 525 Davey Lab, University Park, PA 16802, USA\\
\authormark{4}Penn State Astrobiology Research Center, University Park, PA 16802, USA\\
\authormark{5}Center for Exoplanets \& Habitable Worlds, University Park, PA 16802, USA\\
\authormark{6}Department of Physics and Astronomy, University of Pennsylvania, Philadelphia, PA 19104, USA \\
\authormark{7}NASA Sagan Fellow\\
\authormark{8}jeffrey.m.jennings@colorado.edu\\
\authormark{9}scott.diddams@nist.gov}
    
\begin{abstract}
An optical etalon illuminated by a white light source provides a broadband comb-like spectrum that can be employed as a calibration source for astronomical spectrographs in radial velocity (RV) surveys for extrasolar planets.  For this application the frequency stability of the etalon is critical, as its transmission spectrum is susceptible to frequency fluctuations due to changes in cavity temperature, optical power and input polarization.  In this paper we present a laser frequency comb measurement technique to characterize the frequency stability of a custom-designed fiber {\fabper} interferometer (FFP). Simultaneously probing the stability of two etalon resonance modes, we assess both the absolute stability of the etalon and the long-term stability of the cavity dispersion. We measure mode positions with MHz precision, which corresponds to splitting the FFP resonances by a part in 500 and to RV precision of $\approx 1$ {\ms}. We address limiting systematic effects, including the presence of parasitic etalons, that need to be overcome to push the metrology of this system to the equivalent RV precision of 10 c{\ms}. Our results demonstrate a means to characterize environmentally-driven perturbations of etalon resonance modes across broad spectral bandwidths, as well as motivate the benefits and challenges of FFPs as spectrograph calibrators.
\end{abstract}


\bibliographystyle{osajnl}

\section{Introduction}
Intrinsically stable, broad bandwidth wavelength calibration sources are a necessity in high precision astronomical spectroscopy aimed at measuring Doppler radial velocity (RV) shifts at and below the 1 {\ms} level \cite {Fischer:2016}.  This is equivalent to a fractional Doppler shift of $<3\times 10^{-9}$, or $<2$ femtometers at 600 nm. Such high precision RV measurements are central to the discovery and characterization of exoplanets \cite{Mayor:2003} and could also enable direct measurement of the cosmic expansion \cite{Pasquini:2010}. Conventionally, atomic emission lamps (such as Thorium-Agron and Uranium-Neon) \cite{Mayor:2003, Redman:2011} or molecular absorption cells (I$_2$, CH$_4$) \cite{Butler:1996,Plavchan:2013} have been used as stable wavelength references to calibrate stellar spectra recorded by astronomical spectrographs to a precision of 1 -- 3 {\ms}. These calibrators have been central to the tremendous success of Doppler RV searches for exoplanets over the past decades.  However they also suffer from finite bandwidth, non-uniform spectral features and line blending, making them insufficient for next generation spectrographs aimed at ultimately reaching 1 c{\ms}. 

Recently, laser-based photonic systems have shown promise as calibration tools for precision astronomical spectroscopy \cite{Li:2008, Molaro:2013, Murphy:2007, Philips:2012, Ycas:2012}.  Among these, laser frequency combs (LFCs) have emerged as optimal calibration sources which, when combined with ultra-stable high resolution spectrometers, could enable RV measurement precisions of $<10$ c{\ms} \cite{Probst:2014,Wilken:2012}. A self-referenced optical frequency comb has the unique characteristic of providing a broad array of narrow emission lines whose exact spacing and frequencies are referenced to stabilized frequency standards with fractional uncertainties of $<10^{-12}$ \cite{Quinlan:2010,Ycas:2012}.  Laser frequency combs based on mode-locked lasers and electro-optic frequency modulation have both been employed in proof-of-concept tests at astronomical observatories \cite{Ycas:2012, Philips:2012, Molaro:2013,Kotani:2014,Yi:2016}, and facility-level instruments are presently coming online or being constructed for several state-of-the-art spectrographs under development \cite{Mahadevan:2014a, Pepe:2014b, Kotani:2014, Jurgenson:2016}.

A passive {\fabper} (FP) etalon has also been suggested as an astronomical wavelength calibration source for high precision Doppler measurements \cite{Wildi:2012, Halverson:2014a, Reiners:2014}.  When illuminated by a broadband light source, the transmission of an etalon consists of a broad array of comb-like spectral features. Particular advantages of the FP include rich spectral information content, low optomechanical complexity, and relatively low cost. Our recent work has shown that a fiber-optic integrated implementation, a fiber {\fabper} (FFP), illuminated by a supercontinuum source enabled $<2$ m s$^{-1}$ short-term calibration of the APOGEE near-infrared spectrometer \cite{Halverson:2014a}. However in contrast to broadband optical frequency combs, the FP transmission modes are not strictly uniform in spacing, and their absolute frequencies are tied to the optical and mechanical properties of the etalon itself (rather than a stabilized atomic standard), which can fluctuate and drift with local environmental changes in temperature, pressure, and humidity. Even with precise temperature or mechanical stabilization \cite{Wildi:2012, Halverson:2014a}, or direct optical locking of the FP cavity to an atomic reference \cite{Schwab:2015}, long-term changes of the cavity dispersion and material properties will impact absolute spectral stability of the FP output. Although some research efforts have focused on the use of optical cavities as a frequency transfer tool in atomic spectroscopy and frequency metrology \cite{DeVoe:1988, Banerjee:2003, Singh:2012, Maleki:2010, Jones:2004}, the  ultimate utility of a broadband FP calibrator for astronomy, particularly over long timescales, is not yet fully understood or quantified. 

In this paper we take steps to progress that understanding.  We introduce a technique to characterize the frequency stability of a FP cavity and apply it to a custom FFP that we have constructed for operation in the 780 -- 1350 nm wavelength region.  By simultaneously measuring and tracking multiple FP resonances over days, we can quantify the impact of environmental perturbations on both the absolute stability of the etalon and place an upper limit on changes in the cavity's chromatic dispersion. This technique allows us to identify the line centers of the FP resonance with a precision of $\approx 1$ MHz, equivalent to an RV precision of 1 m s$^{-1}$.  The intrinsic frequency precision of the technique is at the 10 kHz level (RV precision of 1.5 cm s$^{-1}$), and we identify present limitations as arising from uncontrolled temperature gradients, polarization mode dispersion of the FFP and parasitic etalons in the measurement apparatus. The results and the techniques we introduce have significance not only for passively stabilized etalons, but for a variety of actively-stabilized etalon spectral calibration sources that are proposed or are presently being constructed.

\section{Background}
The resonance condition for a {\fabper} interferometer requires that the round trip optical phase shift $k2L$ must be an integer-multiple of $2\pi$,

\begin{equation} 
   2\pi m  = k2L = 2L\omega n/c,
 \end{equation}
where $m$ is an integer, $L$ is the cavity length, $c$ is the speed of light, and $k(\omega)$ and $n(\omega)$ are the frequency-dependent propagation constant and index of refraction, respectively.  This relationship implies that the discrete resonant frequencies $\omega_m$, i.e., the cavity longitudinal modes, are given by 
\begin{equation} 
   \omega_m  = 2\pi m \frac{c} {2Ln} = 2\pi m \times {\Delta}{\nu}_{\rm{FSR}},
 \end{equation}
 where ${\Delta}{\nu}_{\mathrm{FSR}}$ is the cavity's free spectral range.
 
For an ideal dispersionless {\fabper} cavity the frequency response of the resonance modes to changes in the cavity length $\Delta L$ can be illustrated by a simple \lq{}rubber band\rq{} model in which higher frequency modes undergo larger frequency shifts, but for which the fractional frequency shift ($\Delta\omega_m / {\omega_m}$) across the spectrum is constant. The number of modes between two resonance frequencies predicts their relative frequency response to a changing cavity length, analogous to equidistant marks on a rubber band whose separations disperse non-uniformly as the band is stretched. Within this simplified model, the ratio of the frequencies of two modes is independent of $\Delta L$ and given simply by the ratio of their mode numbers, $\omega_m/\omega_l=m/l$. However in a more realistic model the dispersion of the cavity medium must be accounted for.  In this case it is convenient to use a power series expansion of the frequencies of the cavity modes about a nominal central mode $\omega_0$,
 
\begin{equation} 
   \label{eqn:dispersion}
   \omega_m = \omega_0 + D_1m + \frac{1}{2}D_2m^2 + \frac{1}{6}D_3m^3 + ....
 \end{equation}
In this expansion $D_1=2\pi v_g/(2L)$, where the group velocity $v_g=c/(n+\omega(dn/d\omega))$ is evaluated at $\omega_0$. $D_2=-cD_1^2k_2/n$ is related to the group velocity dispersion of light in the cavity with $k_2=d^2k/d\omega^2$, also evaluated at $\omega_0$.  The next term in the expansion accounts for third-order dispersion \cite{Herr:2014}. Here we have also re-indexed the integer $m$ such that its value is zero at $\omega_0$.

In the experiments described in Section~\ref{sec:experimental_setup} we measure the absolute frequency fluctuations of two of the FFP's modes, $m$ and $l$, using probe lasers at 281.6 THz ($\approx$ 1064 nm) and 227.3 THz ($\approx$ 1319 nm). The question then arises: at what magnitude of $\Delta L$ would we expect the measured ratio of frequency shifts, $r=\Delta \omega_m / \Delta \omega_l$ to depart from the constant ratio $m/l \approx 1.239$?  When we include the next higher order ($D_2$) dispersive effects, evaluation of Eq.~\eqref{eqn:dispersion} with parameters of fused silica shows that $\Delta \omega_m / \Delta \omega_l$ departs from a constant ratio by a few parts in $10^{-5}$ for a cavity length variation of $\approx$ half a wavelength (500 nm). This would correspond to an optical frequency shift of one of the FFP modes employed here by $\approx$ 30 GHz, which is a factor of at least 10 greater than any frequency shift we observe.  Thus our analysis implies that if one of the FFP modes could be precisely stabilized with uncertainty $\Delta \omega_l$, the frequency stability of a distant mode could be expected to be similarly stable at the level of $\Delta \omega_m < r\Delta \omega_l(1+10^{-5})$. This deviation is significantly smaller than we can resolve in present experiments.  In Section~\ref{sec:driven_temp} we describe measurement conditions that yield recovery of the expected mode ratio of 1.239 to within a precision as good as $10^{-3}$, limited by our ability to characterize the center frequency of individual FP modes. 

\section{Experimental setup}
\label{sec:experimental_setup}
\begin{figure}
\includegraphics[width=\columnwidth]{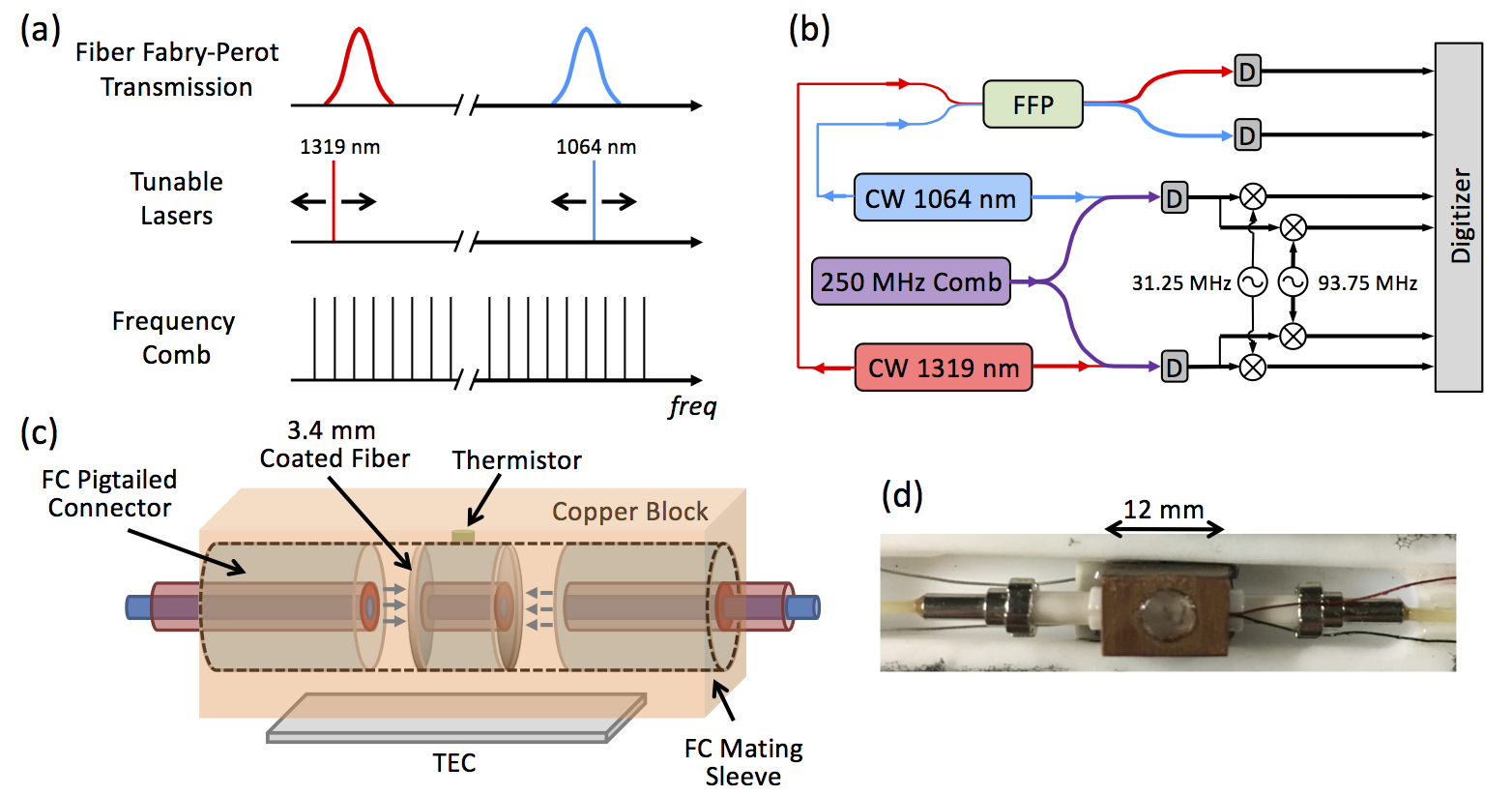}
\caption{(a) Concept of the FFP characterization scheme.  Two CW lasers are scanned across two spectrally displaced FFP resonances while their frequency positions are tracked relative to a 250 MHz, self-referenced optical frequency comb. (b) More detailed experimental configuration showing major optical and RF components used to scan the etalon resonances.  (c) Schematic of the FFP under study, showing thermal heat-sink, thermoelectric cooler (TEC), thermistor, 3.4 mm fiber cavity in a ferrule, and (d) a photo of the same.}
\label{fig:measurement_concept}
\end{figure}

The approach we employ to characterize the stability of the FFP is adapted from that of \cite{DelHaye:2009} and is shown schematically in Fig.~\ref{fig:measurement_concept}.  The concept is to use a near-infrared, self-referenced, octave-spanning LFC \cite{Ycas:2012} to calibrate the scanning of two continuous wave (CW) lasers that are transmitted through the FFP at 281.6 THz and 227.3 THz.  This allows us to simultaneously measure and track the frequencies of two FFP modes in real time.  In principle this approach could be extended to even broader bandwidths by employing additional tunable lasers covering greater spectral range or with the technique of dual-comb frequency spectroscopy \cite{Coddington:2016}.

The FFP has been described in greater detail elsewhere \cite{Halverson:2013}.  As shown in Figs.~\ref{fig:measurement_concept}(c) and \ref{fig:measurement_concept}(d), it consists of an approximately 3.4 mm piece of HI-780 single mode optical fiber that is potted in a ceramic ferrule with reflective coatings on the end faces. The reflective coatings operate over the range of 750 -- 1350 nm with finesse $F$ = 100, and the nominal free spectral range of the etalon is FSR = 30 GHz (0.11 nm at 1064 nm). Additional HI-780 single mode optical fibers are butt-coupled to the FFP to couple light both into and out of it. A 10 k$\Omega$ thermistor sensor is epoxied to the top of the FFP; a 0.5 A thermoelectric cooler (TEC) is epoxied to its central base. These are used with a commercial bench-top temperature controller in a proportional-integral-derivative (PID) loop to maintain a fixed cavity temperature. 

\begin{figure}
\includegraphics[width=\columnwidth]{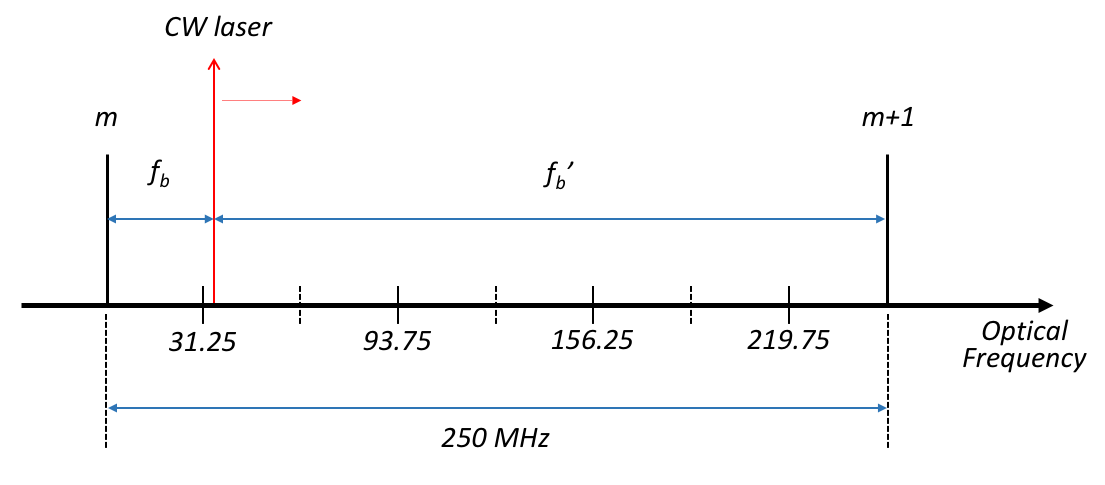}
\caption{Frequency domain diagram of the calibration \lq{}tick\rq{} generation. A CW laser is swept between neighboring frequency comb modes $m$ and $m+1$ of frequency $\nu_m$ and $\nu_{m+1}$, respectively.  This generates two heterodyne beats at frequencies $f_b$ and $f_b'$, which are subsequently mixed against local oscillators at 31.25 MHz and 93.75 MHz.  Within each 250 MHz optical window between adjacent comb modes, four calibration ticks are consequently generated: exactly when $f_b=31.25$ MHz, $f_b=93.75$ MHz, $f_b'=93.75$ MHz, and $f_b'=31.25$ MHz. These occur when the CW laser crosses the points denoted in the diagram at $1/8$, $3/8$, $5/8$, and $7/8$ of the 250 MHz interval.}
\label{fig:tick_generator}
\end{figure}

While the FFP temperature is being controlled, the CW lasers are periodically scanned across their respective FFP resonances by applying a sawtooth tuning voltage profile with period of 100 s to each laser's temperature.  While the start and stop of the scan is synchronous for both lasers, the scan magnitudes are different, and they are offset such that the two wavelengths do not come into resonance simultaneously.  The power incident on the FFP is kept at or below 1 mW in order to minimize laser-induced heating of the FFP that might arise from residual absorption or light scattering.  After transmission the 1064 nm and 1319 nm beams are de-multiplexed in fiber and sent to separate detectors where the transmitted power from each laser is photodetected and digitized. 

Simultaneously, but in two separate detector channels, the CW lasers are heterodyned with the spectrally-broadened output of a self-referenced Er:fiber laser frequency comb (LFC) with repetition rate $f_{\rm rep}=250$ MHz. The LFC is stabilized to a hydrogen maser with absolute optical uncertainty below 100 Hz on the 100 s timescale, which is equivalent to an RV precision of $<0.03$ mm s$^{-1}$. As detailed in Fig.~\ref{fig:tick_generator}, in each $250$ MHz optical window there are two heterodyne beats corresponding to the interference of a CW laser with comb lines $m$ and $m+1$.  As the CW laser is scanned, these heterodyne beats also scan across the 250 MHz window, where they are mixed with RF fixed local oscillators (LOs) at 31.25 MHz and 93.75 MHz, down converted to baseband, low-pass filtered, amplified, and digitized.  Our choice of the LO frequencies provides a calibration \lq{}tick\rq{} each time the CW laser moves 62.5 MHz, or at precisely $f_{\rm rep} / 4$.  We note that our approach of using fixed frequency LOs to down convert to baseband, in contrast to bandpass filters and high-speed digitizers used in \cite{DelHaye:2009}, yields a reduced uncertainty in the frequency calibration.

\begin{figure}
\includegraphics[width=\columnwidth]{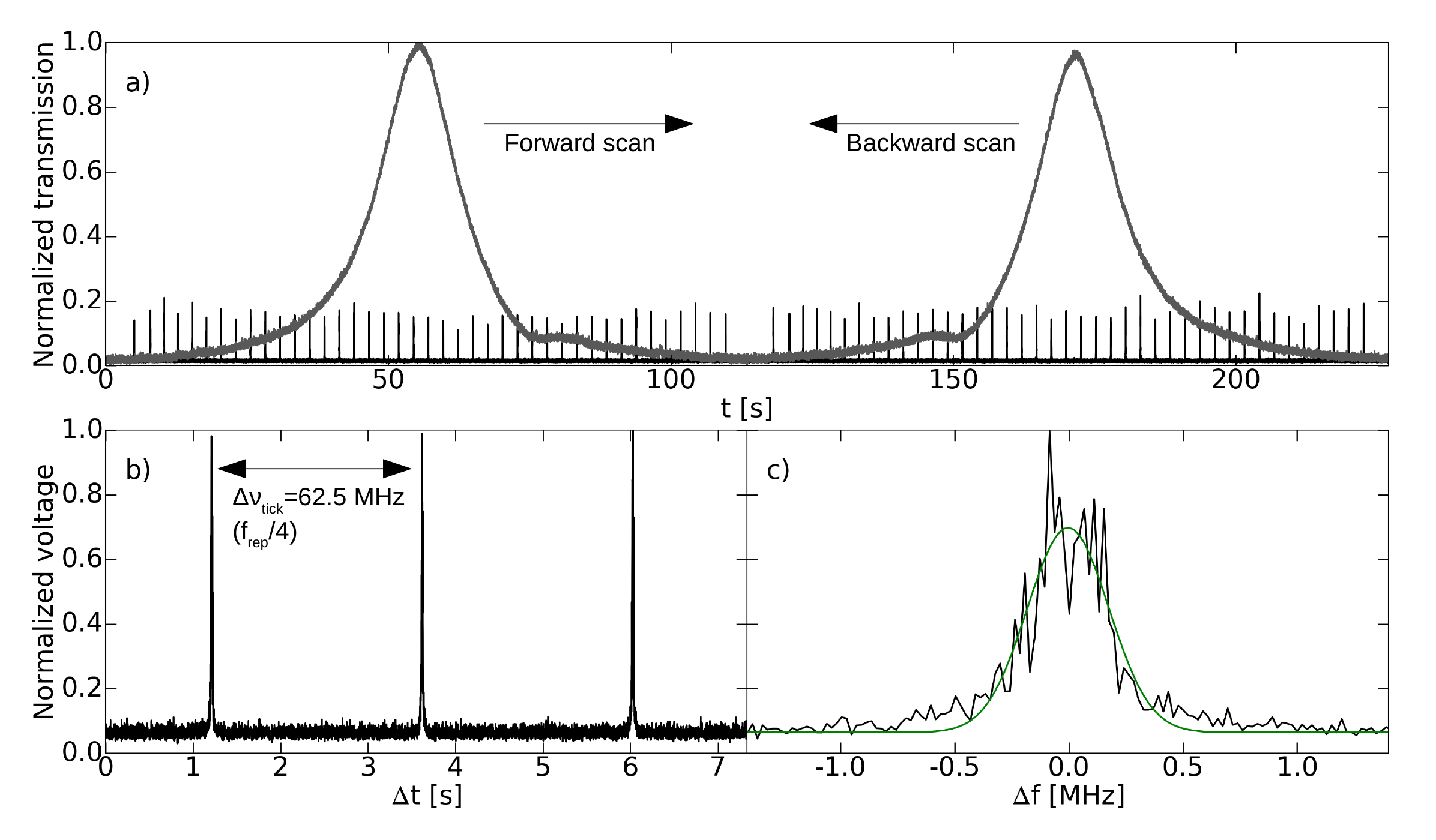}
\caption{(a) A forward (increasing frequency) and backward (decreasing frequency) scan of the laser showing the resonance mode at 1064 nm (grey) and calibration \lq{}ticks\rq{} (black). The gap between the two scans is the read-out period. (b) Zoom on a subset of neighboring frequency ticks. (c) A further zoom on the digitized voltage comprising a single tick and the Gaussian fit that determines its centroid. Here the time acquisition axis has been transformed to a frequency scale (see Fig.~\ref{fig:axis_calibration}) in order to illustrate the achievable precision (this transformation exploits that the spacing is $f_\mathrm{rep} / 4= 62.5$ MHz).} 
\label{fig:scan_example}
\end{figure}

\begin{figure}
\includegraphics[width=\columnwidth]{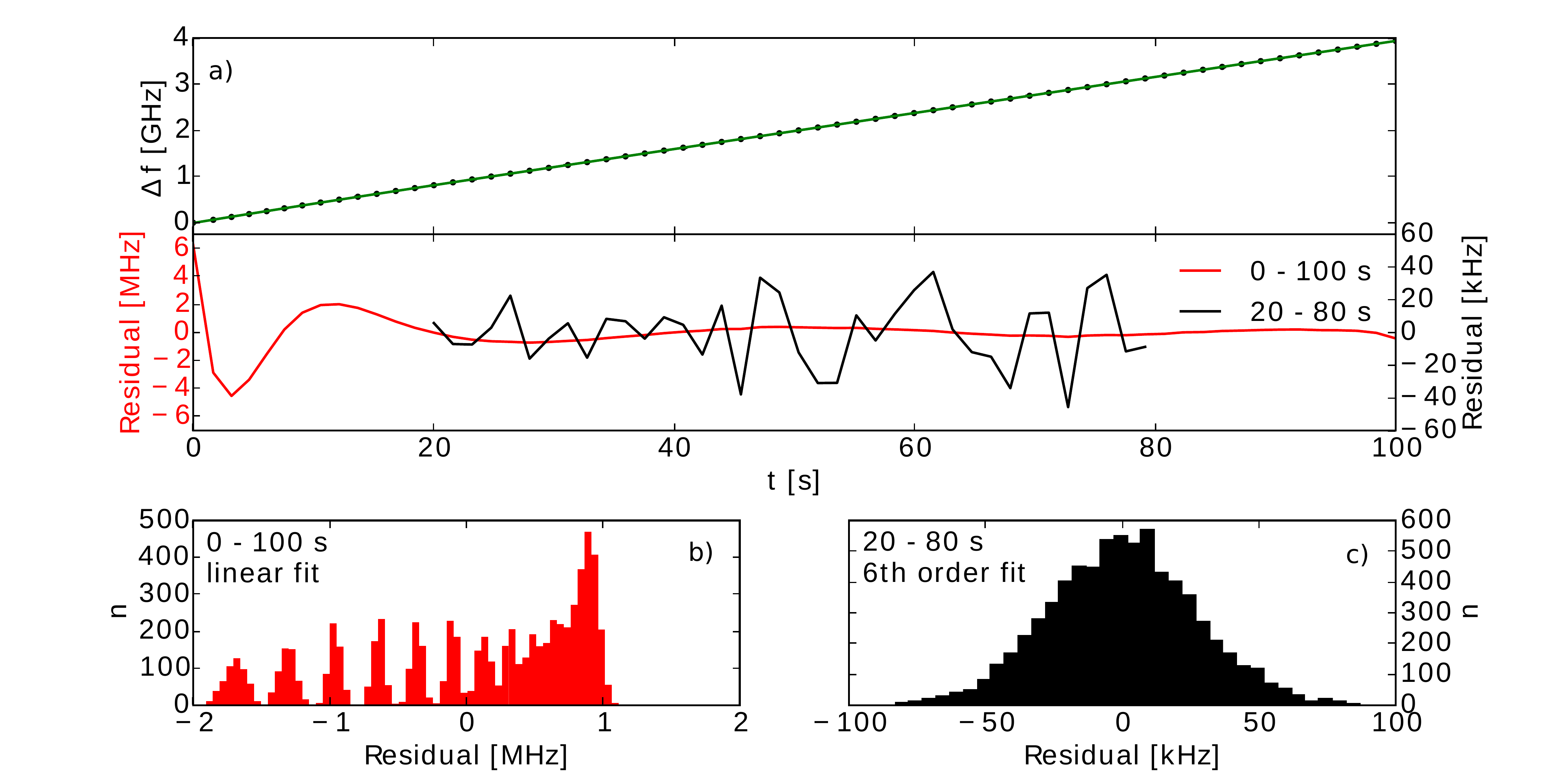}
\caption{(a) Calibration of the frequency axis for a single etalon scan, where each point is the centroid of a Gaussian fit to a calibration tick in Fig.~\ref{fig:scan_example}. Residuals to a linear fit of these points for the full 100 s scan (red, left axis) and a sixth order fit to the central scan region of 20 -- 80 s (black, right axis) are shown in the second panel.  (b) Across $\approx$40 hr of scans, the linear fit to the full scan shows asymmetric residuals. 
(c)  A higher order fit to the central region yields an RMS scatter of 29 kHz.} 
\label{fig:axis_calibration}
\end{figure}

The result of the data acquisition is shown in Fig.~\ref{fig:scan_example}. Once the calibration ticks are digitized we perform a least-squares fit of a Gaussian function to determine the centroid acquisition time (and therefore relative frequency) associated with each tick (see Figs.~\ref{fig:scan_example}(b) and \ref{fig:scan_example}(c)).  The known frequency spacing of the calibration ticks at $f_{\rm rep}/4$ provides pairs of discrete points that map acquisition time to relative frequency. We then use a fit to this data set to remove nonlinearities in the laser frequency scan as shown in Fig.~\ref{fig:axis_calibration}.  A simple linear fit over the full 100 s scan reveals a significant nonlinearity, while a sixth order polynomial fit over the central 60 seconds reduces the uncertainty in our calibration of the frequency axis to $<50$~kHz as indicated by the residuals in Fig.~\ref{fig:axis_calibration}(b).

\begin{figure}
\includegraphics[width=\columnwidth]{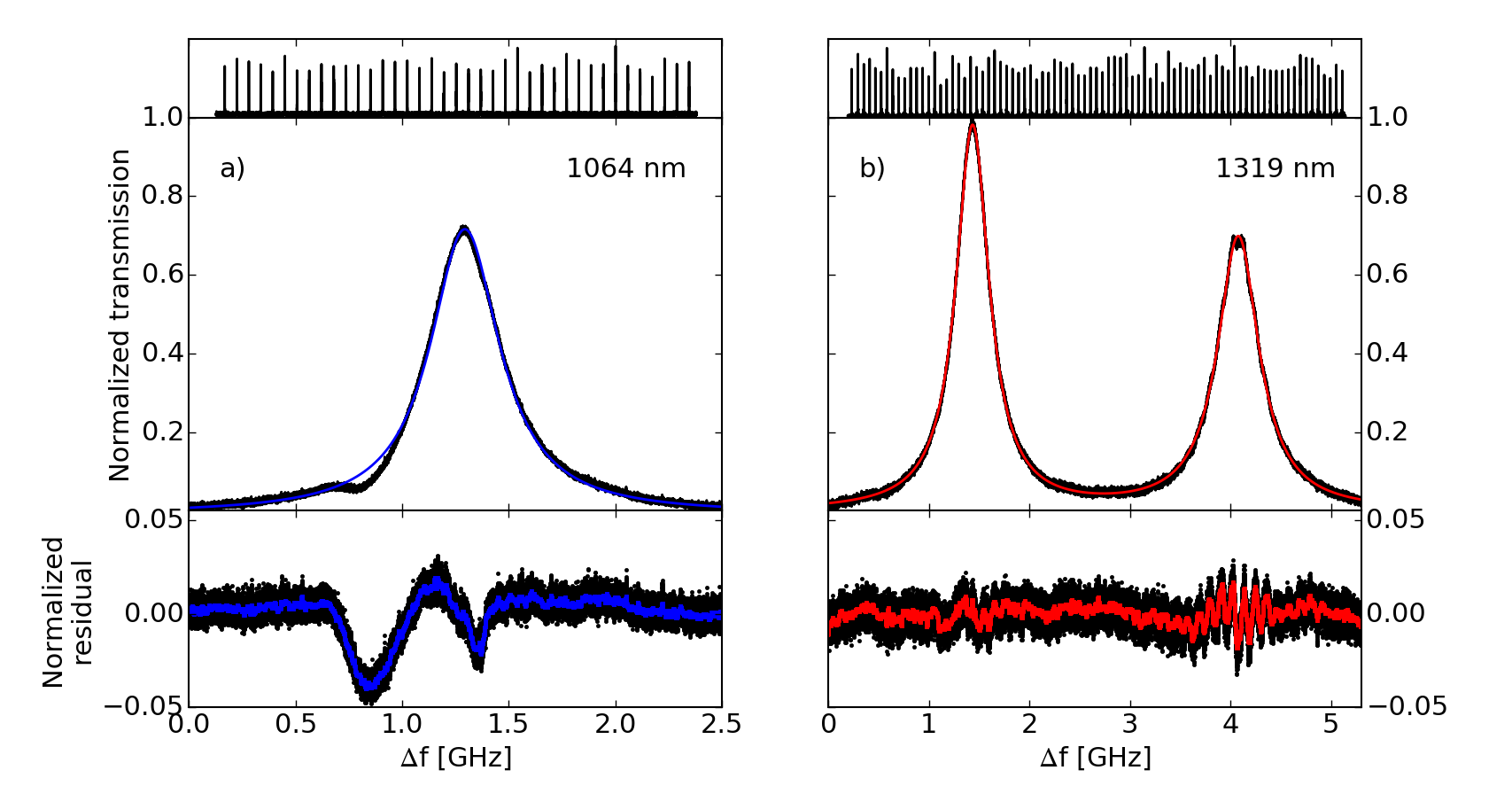}
\caption{A single scan of two etalon resonances using the frequency axis obtained in Fig.~\ref{fig:axis_calibration}, with resonances at (a) 1064 nm and (b) 1319 nm shown with Lorentzian fits and fit residuals. All plots are normalized to the maximum signal voltage of the 1319 nm resonance. A low-pass filter is applied to the residuals, shown in blue and red, respectively. Residuals over the FWHM of the peak at $\Delta f=1.43$ GHz in (b) have an RMS scatter of 0.006. Ticks used to calibrate the frequency axes are shown at top. The asymmetry in the 1064 nm peak is discussed in Section~\ref{sec:experimental_setup}. The effect of parasitic etalons seen in the 1319 nm resonance peaks is discussed in Section~\ref{sec:error_terms}.}
\label{fig:transmission_scan}
\end{figure}

With the frequency axis now determined, the FP transmission resonance data are fit with a Lorentzian (or double Lorentzian, in the case of two prevalent polarization modes of the 1319 nm resonance) profile whose centroid is assigned as the peak resonance frequency. An example case is shown in Fig.~\ref{fig:transmission_scan}. Altering the input polarization in the 1319 nm mode, we can adjust the prominence of the second birefringence peak; it can be brought to parity with the first peak or suppressed into the noise. The frequencies of both polarization modes are tracked across scans, and we observe no difference between them in frequency response. The residuals suggest an achievable noise-limited centroid measurement precision of  $\approx 1$ MHz, limited by the RMS background level of the noise of 0.6\% on a single scan. However these scans also show features in the resonances not described by the Lorentzian fits.  For the 1319 nm resonance we observe higher frequency interference that arises from residual etalons formed between various fiber interfaces in the system.  While we have attempted to minimize these etalons, they still impact the determination of the line center. 

The 1064 nm resonance shows a persistent feature on the low frequency side that we cannot attribute to birefringence or a residual etalon in the external fiber components.  While this discrepancy from the expected Lorentzian transmission profile is stable in time, it too can affect the determination of the line center. We expect to observe two birefringence peaks at every etalon mode (see \cite{Halverson:2014a}). However at 1064 nm we observe no clear changes in the resonance profile (beyond amplitude changes) in response to a varied incident polarization. The profile of the single observed peak is not well described by a combination of two Lorentzians.

To address these effects and verify the accurate tracking of the FFP modes, we independently develop two distinct data fitting and analysis approaches. Both approaches produce results that agree at the 1\% level, but for consistency all numbers quoted here come from a single approach.

\section{Results and discussion}
To illustrate the utility of the techniques outlined above we record several measurement time series, tracking the etalon resonance modes at 1064 and 1319 nm over trials spanning 30 minutes to $>50$ hours. The etalon transmission peaks and heterodyne frequency marker data series are both constantly recorded over the measurement period, and we track the frequencies of the two FP modes while the temperature of the etalon is controlled using two different techniques. In the first (Section~\ref{sec:driven_temp}), we drive the temperature of the etalon with a periodic ramp. In the second (Section~\ref{sec:stable_temp}), we \lq{}fix\rq{} the temperature of the cavity using a tuned PID loop and feedback to the thermoelectric heater/cooler on which the FFP is mounted. The methods yield similar results, though both show structured noise that is likely dominated by systematic effects not directly related to the FFP (discussed in Section~\ref{sec:error_terms}).

\subsection{Applied temperature ramp measurements}
\label{sec:driven_temp}
In all trials, of both the type discussed here and in Section~\ref{sec:stable_temp}, systematic effects (explored in Section~\ref{sec:error_terms}) appear to be biasing our recovered resonance mode frequencies. In an attempt to dominate this bias with a high signal-to-noise, in this set of trials we maximize the relative frequency displacement between the two resonance modes by driving the etalon temperature. We do this by applying a sawtooth voltage profile to the temperature servo setpoint, the resulting low frequency ($1 - 2$ hour periodicity) temperature ramps having amplitude $\gtrsim 0.1$ {\dC} and corresponding temperature gradients between $\approx 0.1 - 4.0$ {\dC} hr$^{-1}$.  These gradients are 1 -- 2 orders of magnitude larger than ambient lab temperature variations and yield a high signal-to-noise measurement of the differential frequency shift between the two resonances. With this method we consistently recover the expected ratio of the 1064 nm and 1319 nm mode frequencies to within measurement precision of $10^{-3}$ fractional. See \cite{Jennings:2016} for a fuller discussion of the thermal response of a single resonance mode over shorter duration trials.  

\begin{figure}
\includegraphics[width=\columnwidth]{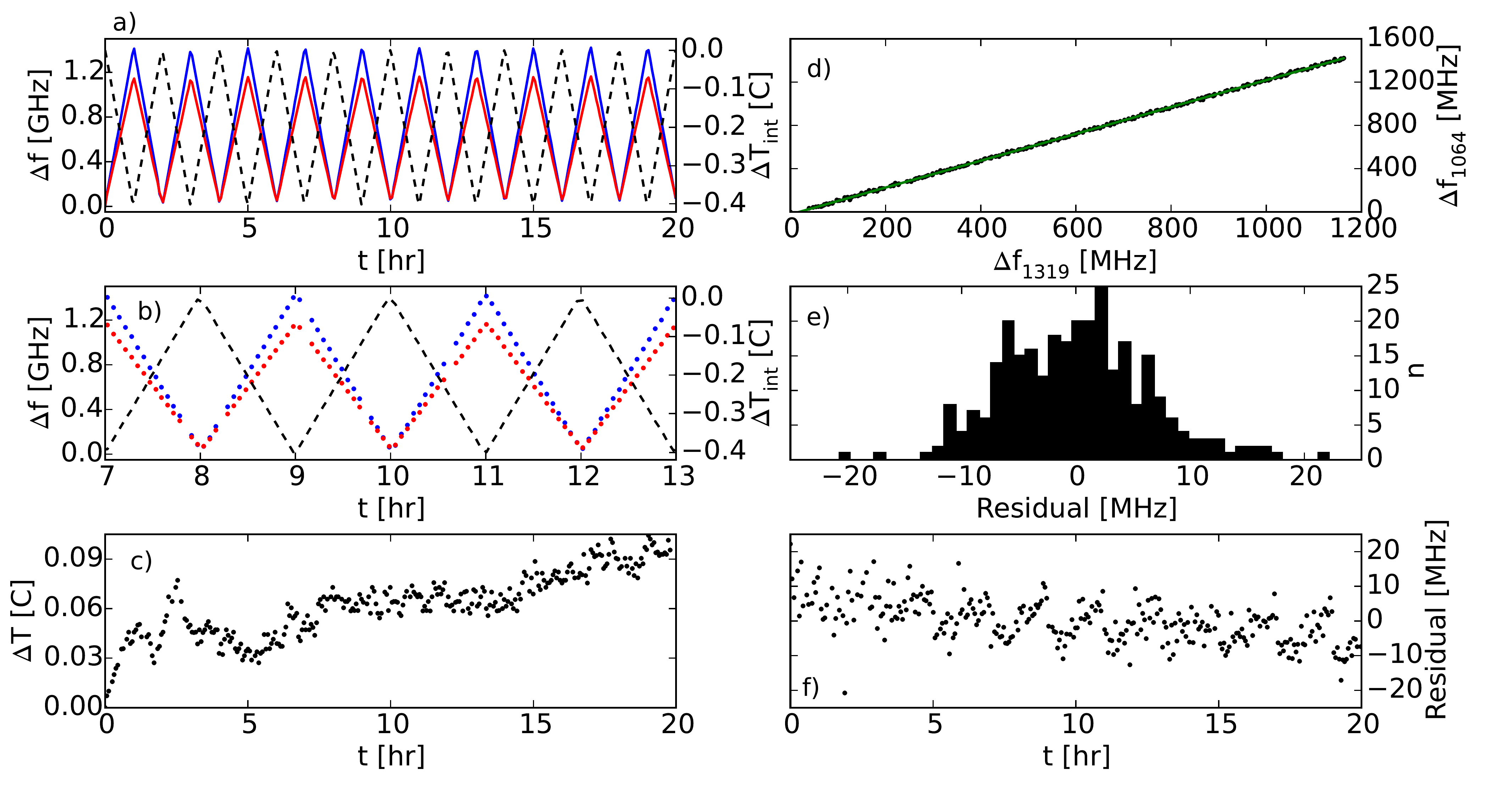}
\caption{(a) A sawtooth voltage ramp is applied to the servo setpoint controlling the cavity's temperature, yielding a temperature gradient of 0.4 {\dC} hr$^{-1}$. The change in the cavity's internal temperature as reported by the servo system, $\Delta T_{\rm int}$ (dashed, black) and corresponding resonance mode responses (blue at 1064 nm, red at 1319 nm) are shown. (b) Zoom on a portion of the scanning period. Gaps occur where the calibrating frequency comb was temporarily unavailable. (c) Change in the ambient lab temperature, $\Delta T$, over the measurement period. (d) Frequency change of the 1064 nm resonance plotted against the drift of the 1319 nm mode, with a linear fit giving a mode ratio of 1.240(0.001). (e) Histogram with 1 MHz bins of the residuals to the linear fit on the data in (d). (f) The fit residuals as a function of time over the measurement period, discussed in Section~\ref{sec:error_terms}.}
\label{fig:driving_ramp}
\end{figure}

Figure~\ref{fig:driving_ramp} shows a 20 hr trial with this measurement technique, using a temperature ramp with amplitude 0.4 {\dC} and 2 hour period. The ratio of the frequency responses at 1064 nm and 1319 nm is described well by a linear fit with slope of $1.240 \pm 0.001$, agreeing within uncertainty with the ratio of the mode numbers. However the residuals to this linear fit show a clear temporal trend (Fig.~\ref{fig:driving_ramp}(f)), this structured noise indicating a temporally varying component to the mode drifts that is not fully quantified by a simple linear fit to the frequency responses of the two modes. The effect is more pronounced when controlling the etalon temperature at a fixed value (Section \ref{sec:stable_temp}) than when deliberately driving with a thermal ramp as done here (discussed further in Section~\ref{sec:error_terms}). 

\subsection{Measurements with FFP temperature control loop engaged}
\label{sec:stable_temp}

\begin{figure}
\includegraphics[width=\columnwidth]{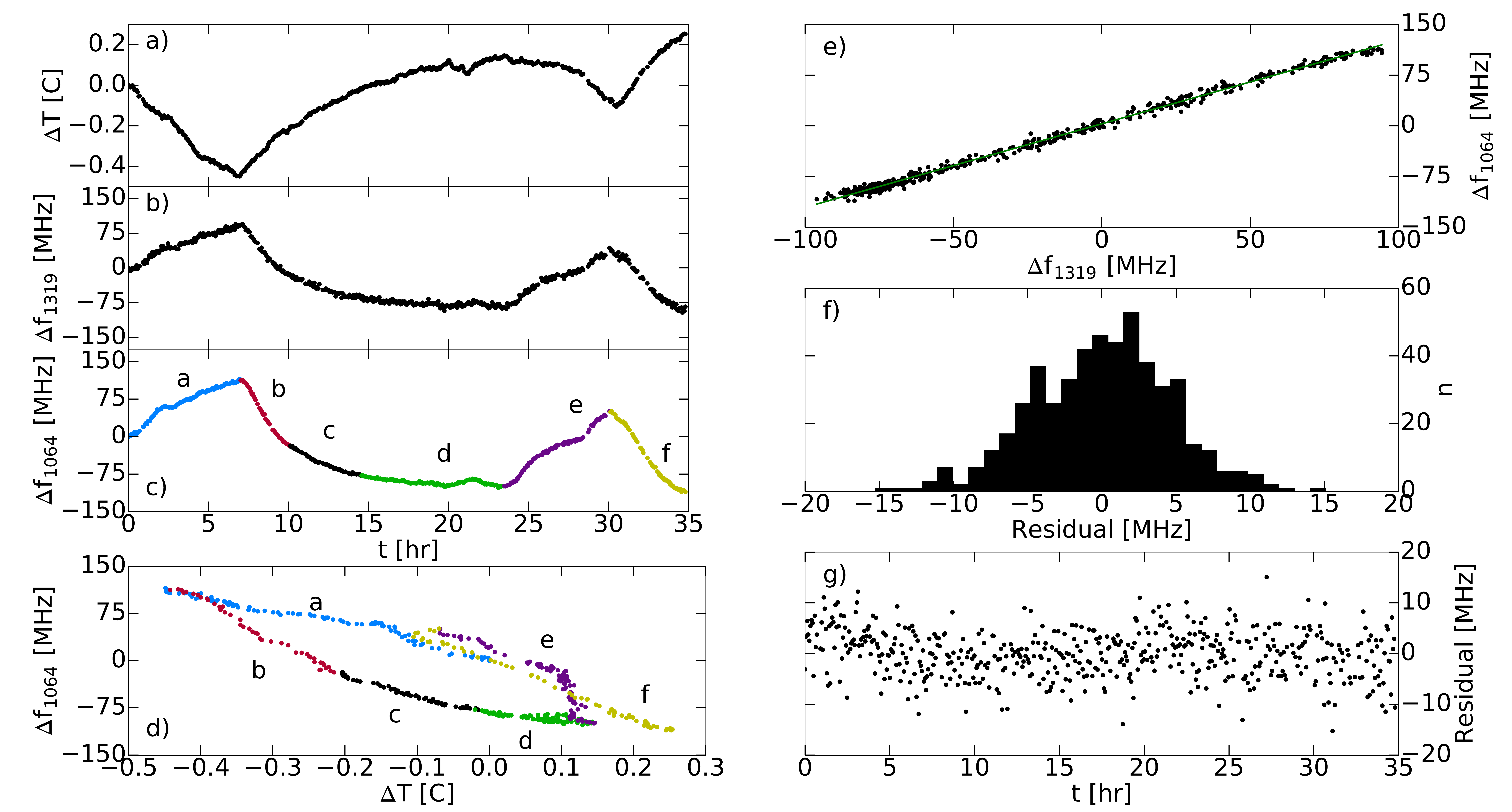}
\caption{Measurement trial for which the FFP thermal control loop is attempting to hold the cavity temperature constant, shown in a similar fashion to Fig.~\ref{fig:driving_ramp}. In spite of the active thermal control loop, a clear correlation is observed between the ambient lab temperature in (a) and the frequencies of the two etalon resonance modes in (b) and (c). Note that the larger range in lab temperatures in (a) compared against Fig.~\ref{fig:driving_ramp}(c) is due to less well regulated $lab$ temperature control during the former, causing the diurnal trend. (d) Frequency of the 1064 nm mode as a function of lab temperature, exhibiting hysteresis. Color codes represent different time segments of the data. (e) Frequency change of the 1064 nm resonance plotted against the drift of the 1319 nm mode, with a linear fit giving a mode ratio of 1.235(0.004). (f) Histogram of the residuals to the linear fit (1 MHz bins). (g) The fit residuals as a function of time over the measurement period, showing measurable structure.}
\label{fig:natural_ramp}
\end{figure}

In the next set of experiments the FFP is controlled at a fixed temperature rather than over an applied periodic temperature ramp, with the consequence in some (but not all) trials being a less accurate measured mode ratio. Consequently, comparison between results with this method and those obtained when driving the cavity temperature rely upon our ability to accurately monitor absolute frequency shifts of the FFP resonances. Figure~\ref{fig:natural_ramp} underscores this point; while the temperature control at the sensing thermistor indicates fluctuations of $<$0.002 {\dC}, we nonetheless observe a strong correlation between the measured mode frequencies and the ambient lab temperature.  This is evidence of insufficient thermal isolation that results in temperature gradients in the present FFP package and is a valuable reminder that in-loop monitoring of such a temperature servo does not guarantee corresponding stability of the FFP. A comparison of Figs.~\ref{fig:natural_ramp}(a)--(c) shows the $<0.5$ {\dC} changes in lab temperature clearly driving the measured resonance mode behavior at 1064 nm and 1319 nm, with higher temperatures shifting the modes to lower frequencies. Focusing on the frequency variations at 1064 nm in Fig.~\ref{fig:natural_ramp}(c), we observe hysteresis when we plot the frequency as a function of lab temperature in Fig.~\ref{fig:natural_ramp}(d). This behavior would likely be reduced with improved thermal isolation. Nonetheless, for this 35 hr trial, we find our measurements of the ratio of the mode frequencies are similar to those presented in Fig.~\ref{fig:driving_ramp}; Figs.~\ref{fig:natural_ramp}(e)--(g) show a mode ratio of 1.235(0.004) over the full run, agreeing within uncertainty with the expected mode ratio. Note that additional trials conducted with this method show agreement with the expected ratio in some cases but departure from it at the $1 \sigma$ level in others. 

\begin{figure}
\includegraphics[width=\columnwidth]{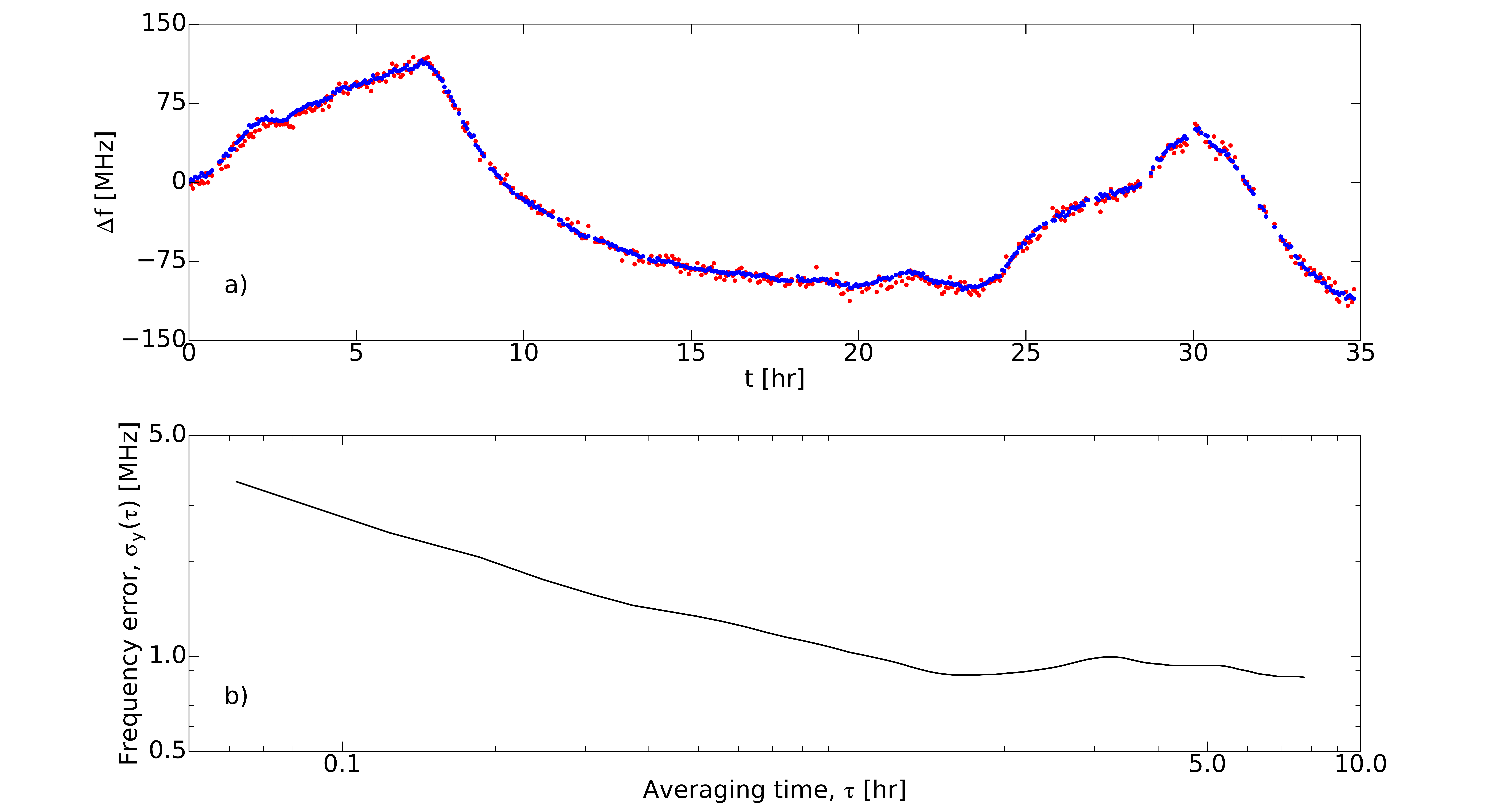}
\caption{(a) Frequency drift ($\Delta f$) of the 1319 nm (red) and 1064 nm modes (blue) for the trial shown in Fig.~\ref{fig:natural_ramp}, where the 1064 nm data have been scaled by the expected mode ratio (1.239) to match the observed frequency shift of the 1319 nm resonance. The 1064 nm and 1319 nm response curves are offset by the difference of their means; in the absence of systematic errors and measurement uncertainties, they should be identical. (b) Allan deviation to the frequency difference in (a). The noise properties of $\Delta f_{1319} - \Delta f_{1064}/1.239$ provide insight into the overall frequency stability of the FFP. See Section~\ref{sec:servo} for further discussion.}
\label{fig:adev}
\end{figure}

\subsection{Simulated frequency lock of the FFP}
\label{sec:servo}
To test the potential for improvement in our stable temperature trial in Figs.~\ref{fig:natural_ramp}(a)--(g), we simulate the case in which the FFP cavity length is locked at one wavelength, e.g., to an optical frequency reference as in \cite{Banerjee:2003,Reiners:2014,Schwab:2015}, with the expectation that frequency stabilizing one mode would stabilize all FFP modes.  The precision with which the out-of-loop modes are stabilized\ --\ particularly those far from the stabilizing frequency\ --\ remains an outstanding question in the field.  In this analysis we use the measured $\Delta f_{1319}$ frequency from Fig.~\ref{fig:natural_ramp} and simulate the effect of a frequency servo with negative feedback by dividing $\Delta f_{1064}$ by $1.239$, the magnitude of the expected mode ratio.  In the ideal case we would expect a 1.239 MHz shift at 1064 nm for a 1 MHz shift at 1319 nm such that $\Delta = \Delta f_{1319} - \Delta f_{1064}/1.239 = 0$. The result, with $\Delta f_{1064}/1.239$ and $\Delta f_{1319}$ overplotted in Fig.~\ref{fig:adev}(a), shows a mean difference of $\Delta = 3.17(0.20)$ MHz (assuming a Gaussian noise distribution). This suggests that if we were to frequency stabilize the 1064 nm mode, limiting factors in our current system would still preclude an accurate prediction of the effect of a perturbation on the 1319 nm mode frequency at the few MHz level. Such limiting factors notably include the larger spread in 1319 nm frequencies relative to 1064 nm in Fig.~\ref{fig:natural_ramp}(a); this may be due to a high frequency parasitic etalon with its own thermal responsivity affecting the 1319 nm measurement (see Fig.~\ref{fig:transmission_scan} and Section~\ref{sec:error_terms}).

The frequency fluctuations in $\Delta$ are shown in the Allan deviation of Fig.~\ref{fig:adev}(b), reaching sub-MHz values for averaging times of $\gtrsim$1 hr. This implies that the actual implementation of a servo could provide frequency stability at the same level. However the flattening of the Allan deviation data at averaging times beyond 1 hour also indicates that the noise processes are no longer Gaussian, as discussed in greater detail below.

\subsection{Factors limiting frequency stability}
\label{sec:error_terms}

We suspect the presence of parasitic etalons, likely induced by low-level reflections at fiber cable connection points within the measurement apparatus, is a limiting factor in the measurement of the intrinsic frequency stability of the FFP. Such parasitic etalons cause the light power transmitted through the FFP to be a combination of a higher frequency modulated signal and the characteristic Lorentzian resonances.  The modulation depends on the temperature and polarization fluctuations in the fiber components external to the FFP itself, but can shift the apparent peak of the Lorentzian resonances. The hypothesis that parasitic etalons are responsible for these behaviors is based on experience with their effects in similar systems, not on a model tested against the data shown here; other potential explanations that may be equally supported by the observations include variable polarization mode coupling between fibers or polarization rotation of the CW lasers. Our interpretation is that due to their unique response functions to temperature, these parasitic etalons gradually shift the fitted resonance frequency of the two modes over time. When the temperature changes are small, the FFP modes themselves are essentially stationary, so the fitted frequency fluctuations are dominated by the parasitic etalons, which cause a systematic bias.  However when the temperature variations are large, the FFP resonances move accordingly and the parasitic etaloning is a smaller perturbation. In trials over which the ambient lab temperature is stable for tens of hours, we see departures from the expected mode ratio as large as the $10^{-2}$ fractional level. If this interpretation is correct it suggests, in agreement with our analytic model in Eq.~\eqref{eqn:dispersion}, that measured departures from the expected mode ratio are a result not of the limited intrinsic frequency stability of the FFP, but instead of imperfections in the optical system. 

\begin{figure}
\includegraphics[width=\columnwidth]{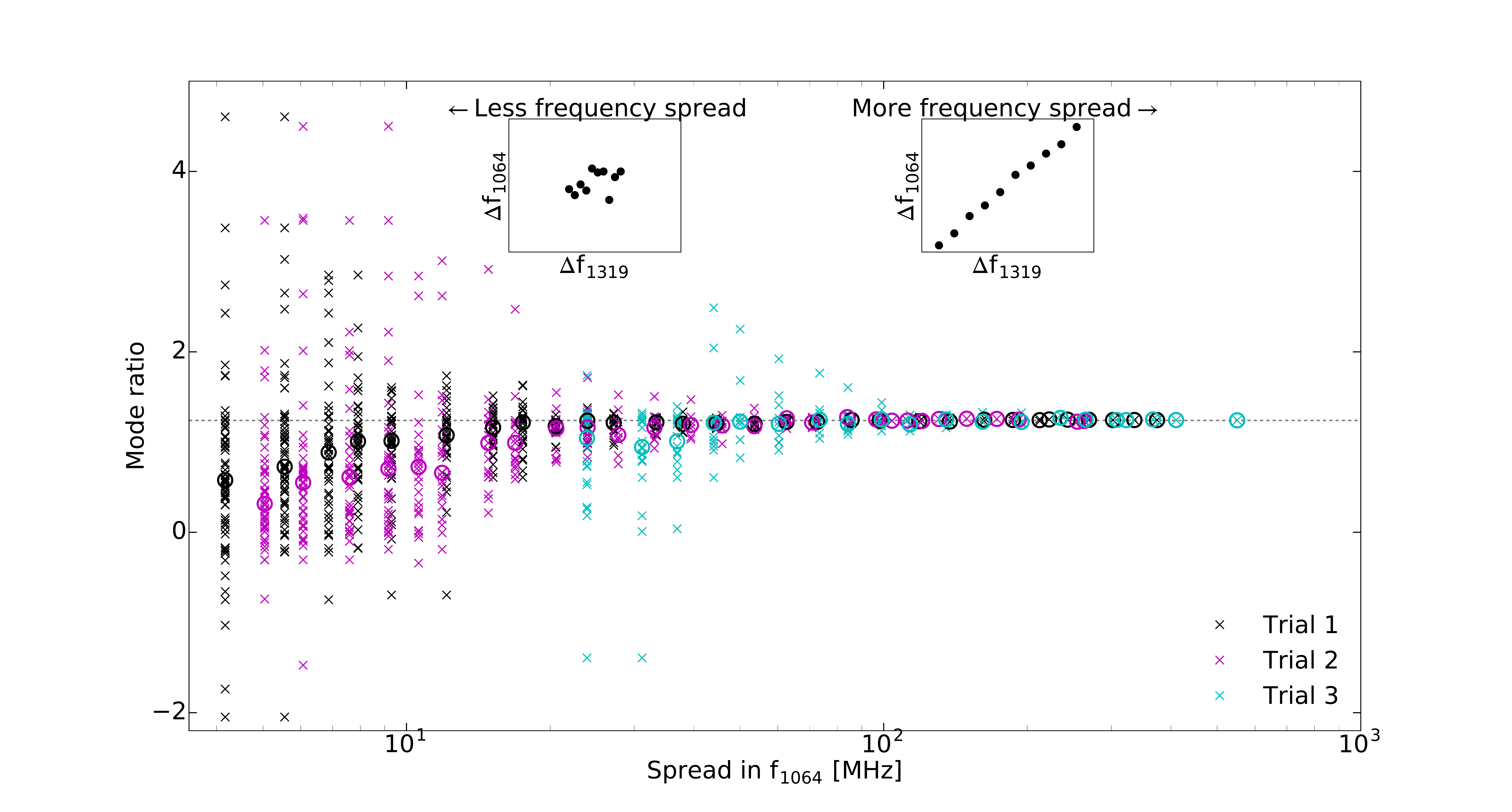}
\caption{Mode ratio as a function of frequency \lq{}spread\rq{} in the 1064 nm resonance mode for three trials in which the FFP temperature is driven in a sawtooth pattern at 0.4 (black crosses; see also Fig.~\ref{fig:driving_ramp}), 0.13 (purple crosses) and 0.08 (blue crosses) {\dC} hr$^{-1}$. Median values are shown as rings. Dashed grey line indicates the expected mode ratio, 1.239. That the expected mode ratio is always approached from below is discussed in Section~\ref{sec:error_terms}. Inset: Qualitative illustration of this $f_{1064}$ spread in the $(\Delta f_{1319},\Delta f_{1064})$ space, with a larger frequency spread corresponding to a convergence on the expected mode ratio.}
\label{fig:mode_ratio_df} 
\end{figure}

We can alternatively frame the mode ratio dependence on an applied temperature gradient as a function of the frequency range (\lq{}spread\rq{}) over which measurements are taken. Temperature gradients of $\lesssim 0.1$ {\dC} hr$^{-1}$ correspond to frequency spreads of $\lesssim 200-300$ MHz, small enough to exhibit departures from the expected mode ratio. These conditions are observed in some trials for which the cavity temperature is held stable. The opposing case, i.e., trials in which the temperature of the etalon is driven significantly, lead to a large frequency shift in both of the probed resonances (and therefore a large difference signal between the two frequencies), and the mode ratio converges to the expected value within measurement uncertainty. This is shown in Fig.~\ref{fig:mode_ratio_df} using data from trials in which the cavity temperature is driven. Here the mode ratio is determined by taking successively larger bins in $\Delta f_{1064}$, with bins that span a larger frequency spread showing a mode ratio closer to the expected value.

For bins in these driven temperature trial datasets with a low frequency spread (which would be equivalent to measurement trials with a nearly constant temperature), the mode ratio is systematically lower than predicted. A possible explanation for this bias in the mode ratio measurement is the thermal response of high frequency parasitic etalons present in the 1319 nm mode measurements (see Fig.~\ref{fig:transmission_scan}). Note that all measurements, whether the cavity temperature is driven or held stable, are likely susceptible to this error source. Its effect may in the stable temperature case be an increased scatter in measured $\Delta f_{1319}$ frequencies, the uncertainty in the mode ratio measured in Fig.~\ref{fig:natural_ramp} (the most accurate of our stable temperature trials in recovering the expected mode ratio) being a factor of 4 larger than the driven temperature case of Fig.~\ref{fig:driving_ramp}. The increased spread in 1319 nm relative to 1064 nm mode frequencies is seen further in Fig.~\ref{fig:adev}(a). Moreover, as mentioned prior, some additional trials not shown here in which the cavity temperature is stabilized show departures from the expected ratio at the $1\sigma$ level. 

The effect of these parasitic etalons in the driven temperature trials is suppressed by improved signal-to-noise but still apparent. The slow temporal drift in the residuals to a linear fit on the frequency response of the two resonance modes in Fig.~\ref{fig:driving_ramp}(f) may be the result of parasitic etalons gradually pulling our fit for one or both modes in a preferential direction. This temporal structure may alternatively suggest the inadequacy of our explanatory model to separate the effects of the parasitic etalons and FFP resonances. To this end we have tested the effect of fitting out the largest amplitude parasitic etalons, but see no significant deviation from fitted peak frequencies obtained in individual scans.

Finally, we additionally test resonance mode sensitivities to variations in incident power. Instantaneous increases in power (contributed equally by the two CW sources) sent into the etalon between factors of $\approx$2 -- 10 (at most, a step from 100 $\mu$W -- 1 mW) show no clear effect on the mode ratio. This agrees with our previous results; by spiking input power we introduce a strong temperature gradient, to which the mode ratio responds according to the theoretical expectation.

\subsection{Steps to improve performance}
\label{sec:error_sources}
Based on our findings we suggest the following to improve FFP frequency stability:

i) minimize parasitic etalons. Parasitic etalons seem to pull resonance modes in a complex and dynamic manner. Fusion splicing the entire optical path up to the etalon would be the optimal solution.

ii) maintain polarization throughout the system; reduce polarization sensitivity. Variations in incident polarization significantly alter the relative amplitude of the two birefringence peaks in our 1319 nm mode; a peak whose frequency is tracked may in the worst case be reduced to parity with the transmission baseline under polarization changes.

iii) construct the etalon with high heat capacity materials. The use of, e.g., silicon ferrules would provide high thermal conductivity and a low coefficient of thermal expansion, improving the FFP's thermal response.

\section{Conclusions}
We have presented a measurement scheme to assess the frequency stability of a fiber {\fabper} etalon, monitoring the relative response of two resonance modes to temperature and properties of the optical system. We observed a trend toward higher uncertainty for measurements taken over a smaller frequency spread (or equivalently, smaller temperature range), which we suspect is due to the presence of parasitic etalons rather than an inherent limitation of the device. We offered a simple physical model for the FFP's dispersion and considered factors in the experimental setup that complicate this behavior, offering steps to mitigate adverse effects. 

In its present form the FFP can provide a stable short-term frequency reference.  However over a longer term (hours to days), one mode of the FFP would have to stabilized against an absolute reference.  With such stabilization, our results predict that the frequencies of additional FFP modes are simply related by the ratio of mode numbers, with a precision as good as $\approx 1$ MHz ($\approx 1$ m s$^{-1}$ equivalent RV precision).  While the FFP may not approach the precision levels seen with atomically-stabilized laser frequency combs, the simplicity of the design remains appealing, and with careful treatment of systematic effects described here, the performance of an FFP for precision RV spectroscopy can likely be improved to the level of 10 cm s$^{-1}$.

\section*{Funding}
The Center for Exoplanets and Habitable Worlds in the Eberly College of Science of the Pennsylvania State University and the Pennsylvania Space Grant Consortium; National Science Foundation (NSF) (AST 1310875, AST 1006676, AST 1126413, AST 1310885); NASA Astrobiology Institute (NNA09DA76A).

\section*{Acknowledgments}
This work was performed in part under contract with the 
California Institute of Technology/Jet Propulsion Laboratory funded by NASA through the Sagan Fellowship Program executed by the NASA Exoplanet Science Institute.  We thank Andrew Metcalf, Scott Papp and Franklyn Quinlan for their helpful comments on this manuscript. We are grateful to the anonymous reviewers for their thoughtful and diligent suggestions. JJ thanks S. Jennings for her contributions to the work. This work is a contribution of NIST and is not subject to copyright in the US.


\begin{thebibliography}{21}
\bibitem{Fischer:2016}
 D.~A. Fischer, G. Anglada-Escude, P. Arriagada, R.~V. Baluev, J.~L. Bean, F.  Bouchy, L.~A. Buchhave, T. Carroll, A. Chakraborty, J.~R. Crepp, R.~I. Dawson, S. Diddams, X. Dumusque, J.~D. Eastman, M. Endl, P. Figueira, E.~B. Ford, D. Foreman-Mackey, P. Fournier, G. F{\H u}r{\'e}sz, B.~S. Gaudi, P.~C. Gregory, F. Grundahl, A.~P. Hatzes, G. H{\'e}brard, E. Herrero, D.~W. Hogg, A.~W. Howard, J.~A. Johnson, P. Jorden, C.~A. Jurgenson, D.~W. Latham, G. Laughlin, T.~J. Loredo, C. Lovis, S. Mahadevan, T.~M. McCracken, F. Pepe, M. Perez, D.~F. Phillips, P.~P. Plavchan, L. Prato, A. Quirrenbach, A. Reiners, P. Robertson, N.~C. Santos, D. Sawyer, D. Segransan, A. Sozzetti, T. Steinmetz, A. Szentgyorgyi, S. Udry, J.~A. Valenti, S.~X. Wang, R.~A. Wittenmyer, and J.~T. Wright, ``State of the field: extreme precision radial velocities,'' Publications of the Astronomical Society of the Pacific {\bf 128}(964), 066001 (2016).

 \bibitem{Mayor:2003}
  M. Mayor, F. Pepe, D. Queloz, F. Bouchy,
   G. Rupprecht, G. Lo Curto, G. Avila, W. Benz, J.~L. Bertaux, X. Bonfils, Th. Dall, H. Dekker, B. Delabre, W. Eckert, M. Fleury, A. Gilliotte, D. Gojak, J.~C. Guzman, D. Kohler, J.~L. Lizon, A. Longinotti, C. Lovis, D. Megevand, L. Pasquini, J. Reyes, J.~P. Sivan, D. Sosnowska, R. Soto, S. Udry, A. van Kesteren, L. Weber, and U. Weilenmann, ``Setting new standards with HARPS,'' The Messenger {\bf 114}, 20--24 (2003).
   
 \bibitem{Pasquini:2010}
  L. Pasquini, S. Cristiani, R. Garcia-Lopez,
   M. Haehnelt, and M. Mayor, ``CODEX: An ultra-stable high resolution spectrograph for the E-ELT,'' The Messenger {\bf 140}, 20--21 (2010).

 \bibitem{Redman:2011}
  S.~L. Redman, J.~E. Lawler, G. Nave, L.~W. Ramsey, \&
   S. Mahadevan, ``The infrared spectrum of uranium hollow cathode lamps from 850 nm to 4000 nm: wavenumbers and line Identifications from Fourier transform spectra,'' The Astrophysical Journal Supplement Series {\bf 195}, 24--33 (2011).
   
 \bibitem{Butler:1996}
 R.~P. Butler, G.~W. Marcy, E. Williams, C. McCarthy, P. Dosanjh,
   and S.~S. Vogt, ``Attaining Doppler precision of 3 m s$^{-1}$,'' Publications of the Astronomical Society of the Pacific {\bf 108}, 500--509 (1996).

 \bibitem{Plavchan:2013}
  P.~P. Plavchan, G. Anglada-Escude, R. White, P. Gao, C. Davison, S. Mills, C. Beichman, C. Brinkworth, J. Johnson, M. Bottom, D. Ciardi, K. Wallace, B. Mennesson, K. von Braun, G. Vasisht, L. Prato, S. Kane, A. Tanner, B. Walp, S. Crawford, and S. Lin, ``Precision near-infrared radial velocity instrumentation I: absorption gas cells,'' Proc.\ Soc.\ Photo-Opt.\ Instrum.\ Eng.\ {\bf 8864}, 88641J (2013).
 
 \bibitem{Li:2008}
  C.~H. Li, A.~J. Benedick, P. Fendel, A.~G. Glenday,
   F.~X. K\"artner, D.~F. Phillips, D. Sasselov, A. Szentgyorgyi, and R.~L. Walsworth, ``A laser frequency comb that enables radial velocity measurements with a precision of 1 cm s$^{-1}$,'' Nature {\bf 452}, 610--612 (2008).
   
 \bibitem{Molaro:2013}
  P. Molaro, M. Esposito, S. Monai, G. Lo Curto,
   J.~I. Gonz\'alez Hern\'andez, T.~W. H\"ansch, R. Holzwarth, A. Manescau, L. Pasquini, R.~A. Probst, R. Rebolo, T. Steinmetz, Th. Udem, and T. Wilken, ``A frequency comb calibrated solar atlas,'' Astronomy \& Astrophysics {\bf 560}, A {\bf 61} (2013).  
 
 \bibitem{Murphy:2007}
  M.~T. Murphy, T. Udem, R. Holzwarth, A. Sizmann,
   L. Pasquini, C. Araujo-Hauck, H. Dekker, S. D'Odorico, M. Fischer, T.~W. H{\"a}nsch, and A. Manescau, ``High-precision wavelength calibration of astronomical spectrographs with laser frequency combs,'' Monthly Notices of the Royal Astronomical Society {\bf 380}(2), 839--847 (2007).
 
 \bibitem{Philips:2012} 
  D.~F. Phillips, A.~G. Glenday, C.~H. Li, C. Cramer,
   G. Furesz, G. Chang, A.~J. Benedick, L. Chen, F.~X. K{\"a}rnter, S. Korzennik, D. Sasselov, A. Szentgyorgyi, and R.~L. Walsworth, ``Calibration of an astrophysical spectrograph below 1 m/s using a laser frequency comb,'' Opt.\ Express {\bf 20}(13), 13711--13726 (2012).
   
 \bibitem{Ycas:2012}
  G.~G. Ycas, F. Quinlan, S.~A. Diddams, S. Osterman,
   S. Mahadevan, S. Redman, R. Terrien, L. Ramsey, C.~F. Bender, B. Botzer, and S. Sigurdsson, ``Demonstration of on-sky calibration of astronomical spectra using a 25 GHz near-IR laser frequency comb,'' Opt.\ Express {\bf 20}(6), 6631--6643 (2012).

 \bibitem{Coddington:2016}
  I. Coddington, N. Newbury, and W. Swann, ``Dual-comb spectroscopy,'' Optica {\bf 3}(4), 414--426 (2016).
  
 \bibitem{Probst:2014}
  R.~A. Probst, G. Lo~Curto, G. Avila, B.~L. Canto~Martins,
   J.~R. de~Medeiros, M. Esposito,  J.~I. Gonz{\'a}lez Hern{\'a}ndez, T.~W. H{\"a}nsch, R. Holzwarth, F. Kerber, I.~C. Le{\~a}o, A. Manescau, L. Pasquini, R. Rebolo-L{\'o}pez, T. Steinmetz, T. Udem, and Y. Wu, ``A laser frequency comb featuring sub-cm/s precision for routine operation on HARPS,'' Proc.\ Soc.\ Photo-Opt.\ Instrum.\ Eng.\  {\bf 9147}, 91471C (2014).

 \bibitem{Wilken:2012}
  T. Wilken, G. Lo~Curto, R.~A. Probst,	T. Steinmetz, A. Manescau, L. Pasquini, J.~I. Gonz{\'a}lez Hern{\'a}ndez, R. Rebolo, T.~W. H{\"a}nsch, T. Udem, and R. Holzwarth, ``A spectrograph for exoplanet observations calibrated at the centimetre-per-second level,'' Nature {\bf 485}, 611--614 (2012).
  
 \bibitem{Quinlan:2010}
F. Quinlan, G.~G. Ycas, S. Osterman, and S.~A. Diddams, ``A 12.5 GHz-spaced optical frequency comb spanning >400 nm for near-infrared astronomical spectrograph calibration,'' Review of Scientific Instruments {\bf 81}, 063105 (2010).

 \bibitem{Kotani:2014}
  T. Kotani M. Tamura, H. Suto, J. Nishikawa, B. Sato, W. Aoki, T. Usuda, T. Kurokawa, K. Kashiwagi, S. Nishiyama, Y. Ikeda, D.~B. Hall, K.~W. Hodapp, J. Hashimoto, J.-I. Morino, Y. Okuyama, Y. Tanaka, S. Suzuki, S. Inoue, J. Kwon, T. Suenaga, D. Oh, H. Baba, N. Narita, E. Kokubo, Y. Hayano, H. Izumiura, E. Kambe, T. Kudo, N. Kusakabe, M. Ikoma, Y. Hori, M. Omiya, H. Genda, A. Fukui, Y. Fujii, O. Guyon, H. Harakawa, M. Hayashi, M. Hidai, T. Hirano, M. Kuzuhara, M. Machida, T. Matsuo, T. Nagata, H. Onuki, M. Ogihara, H. Takami, N. Takato, Y.~H. Takahashi, C. Tachinami, H. Terada, H. Kawahara, and T. Yamamuro, ``Infrared Doppler instrument (IRD) for the Subaru telescope to search for Earth-like planets around nearby M-dwarfs,'' Proc.\ Soc.\ Photo-Opt.\ Instrum.\ Eng.\ {\bf 9147}, 914714 (2014).
 
 \bibitem{Yi:2016}
  X. Yi, K. Vahala,	J. Li,	S. Diddams,	G. Ycas, P. Plavchan, S. Leifer, J. Sandhu, G. Vasisht, P. Chen, P. Gao, J. Gagne, E. Furlan, M. Bottom, E.~C. Martin, M.~P. Fitzgerald, G. Doppmann, and C. Beichman, ``Demonstration of a near-IR line-referenced electro-optical laser frequency comb for precision radial velocity measurements in astronomy,'' Nature Communications {\bf 7}, 10436 (2016).

  \bibitem{Mahadevan:2014a}
  S. Mahadevan L.~W Ramsey, R. Terrien,
   S.~P. Halverson, A. Roy, F. Hearty, E. Levi, G.~K. Stefansson, P. Robertson, C. Bender, C. Schwab, and M. Nelson, ``The Habitable-zone Planet Finder: a status update on the development of a stabilized fiber-fed near-infrared spectrograph for the for the Hobby-Eberly telescope,'' Proc.\ Soc.\ Photo-Opt.\ Instrum.\ Eng.\ {\bf 9147}, 91471G (2014).
 
 \bibitem{Pepe:2014b}
  F. Pepe, P. Molaro, S. Cristiani,
   R. Rebolo, N.~C. Santos, H. Dekker, D. M{\'e}gevand, F.~M. Zerbi, A. Cabral, P. Di Marcantonio, M. Abreu, M. Affolter, M. Aliverti, C. Allende Prieto, M. Amate, G. Avila, V. Baldini, P. Bristow, C. Broeg, R. Cirami, J. Coelho, P. Conconi, I. Coretti, G. Cupani, V. D'Odorico, V. De Caprio, B. Delabre, R. Dorn, P. Figueira, A. Fragoso, S. Galeotta, L. Genolet, R. Gomes, J.~I. Gonz{\'a}lez Hern{\'a}ndez, I. Hughes, O. Iwert, F. Kerber, M. Landoni, J.-L. Lizon, C. Lovis, C. Maire, M. Mannetta, C. Martins, M. Monteiro, A. Oliveira, E. Poretti, J.~L. Rasilla, M. Riva, S. Santana Tschudi, P. Santos, D. Sosnowska, S. Sousa, P. Span{\'o}, F. Tenegi, G. Toso, E. Vanzella, M. Viel, and M.~R. Zapatero Osorio, ``ESPRESSO: the next European exoplanet hunter,'' Astronomische Nachrichten {\bf 335}(1), 8--20 (2014).
 
 \bibitem{Jurgenson:2016}
  C. Jurgenson, D. Fischer, T. McCracken, D. Sawyer, A. Szymkowiak, A.~B. Davis, G. Muller, and F. Santoro, ``EXPRES: a next generation RV spectrograph in the search for Earth-like worlds,'' Proc.\ Soc.\ Photo-Opt.\ Instrum.\ Eng.\ {\bf 9908}, 99086T (2016).

 \bibitem{Wildi:2012}
  F. Wildi, B. Chazelas, and F. Pepe, ``A passive cost-effective solution for the high accuracy wavelength calibration of radial velocity spectrographs,'' Proc.\ Soc.\ Photo-Opt.\ Instrum.\ Eng.\ {\bf 8446}, 84468E (2012).
  
 \bibitem{Halverson:2014a}
  S.~P. Halverson, S. Mahadevan, L. Ramsey, F. Hearty, J. Wilson, J. Holtzman, S. Redman, G. Nave, D. Nidever, and M. Nelson, ``Development of fiber Fabry-P{\'e}rot interferometers as stable near-infrared calibration sources for high resolution spectrographs,'' Publications of the Astronomical Society of the Pacific {\bf 126}(939), 445--458 (2014).  
 
 \bibitem{Reiners:2014}
   A. Reiners, R.~K. Banyal and R.~G. Ulbrich, ``A laser-lock concept to reach cm s$^{-1}$-precision in Doppler experiments with Fabry-P{\'e}rot wavelength calibrators,'' Astronomy \& Astrophysics {\bf 569}, A77 (2014).
   
 \bibitem{Schwab:2015}
  C. Schwab, J. St\"urmer, Y.~V. Gurevich,
   T. F\"uhrer, S.~K. Lamoreaux, T. Walther, and A. Quirrenbach, ``Stabilizing a Fabry-P{\'e}rot etalon peak to 3 cm s$^{-1}$ for spectrograph calibration,'' Publications of the Astronomical Society of the Pacific {\bf 127}(955), 880--889 (2015).
   
 \bibitem{DeVoe:1988}
  R.~G. DeVoe, C. Fabre, K. Jungmann, J. Hoffnagle, and R.~G. Brewer, ``Precision optical-frequency-difference measurements,'' Phys.\ Rev.\ A {\bf 37}, 1802--1805 (1988).  
 
 \bibitem{Banerjee:2003}
  A. Banerjee, D. Das, and V. Natarajan, ``Precise frequency measurements of atomic transitions by use of a Rb-stabilized resonator,'' Opt.\ Lett.\ {\bf 28}(17), 1579--1581 (2003). 
  
 \bibitem{Singh:2012}
  A.~K. Singh, L. Muanzuala, A.~K. Mohanty, and V. Natarajan, ``Optical frequency metrology with an Rb-stabilized ring-cavity resonator -- study of cavity-dispersion errors,'' J.\ Opt.\ Soc.\ Am.\ B {\bf 29}(10), 2734--2740 (2012).  

 \bibitem{Maleki:2010}
  L. Maleki, V.~S. Ilchenko, M. Mohageg, A.~B. Matsko, A.~A. Savchenkov, D. Seidel, N.~P. Wells, J.~C. Camparo, and B. Jaduszliwer, ``All-optical integrated atomic clock,'' 2010 IEEE International Frequency Control Symposium, 119--124 (2010).
  
   \bibitem{Jones:2004}
  R.~J. Jones, I. Thomann, and J. Ye, ``Precision stabilization of femtosecond lasers to high-finesse optical cavities,'' Phys.\ Rev.\ A {\bf 69}, 051803(R) (2004).
  
 \bibitem{Herr:2014}
  T. Herr, V. Brasch, J.~D. Jost, I. Mirgorodskiy, G. Lihachev, M.~L. Gorodetsky, and  T.~J. Kippenberg, ``Mode spectrum and temporal soliton formation in optical microresonators,'' Phys.\ Rev.\ Lett.\ {\bf 113}(12), 123901 (2014).

\bibitem{DelHaye:2009}
 P. Del'Haye, O. Arcizet, M.~L. Gorodetsky, R. Holzwarth, \&
   T.~J. Kippenberg, ``Frequency comb assisted diode laser spectroscopy for measurement of microcavity dispersion,'' Nat.\ Photon.\ {\bf 3}, 529--533 (2009).

 \bibitem{Halverson:2013}
  S.~P. Halverson, S. Mahadevan, and L. Ramsey, ``A fiber Fabry-P{\'e}rot interferometer as stable wavelength reference for high-resolution astronomical spectrographs,'' Workshop on Specialty Optical Fibers and their Applications (Optical Society of America), W3.3 (2013).

 \bibitem{Jennings:2016}
  J. Jennings, S.~P. Halverson, S.~A. Diddams, R. Terrien, G.~G. Ycas, and S. Mahadevan, ``Measuring the thermal sensitivity of a fiber Fabry-P{\'e}rot interferometer,'' Proc.\ Soc.\ Photo-Opt.\ Instrum.\ Eng.\ {\bf 9907}, 99072G (2016).

 \end{thebibliography}
\end{document}